# COOLING GAS OUTFLOWS FROM GALAXIES

Boqi Wang

Department of Physics and Astronomy
The Johns Hopkins University, Baltimore, MD 21218

# COOLING GAS OUTFLOWS FROM GALAXIES

Boqi Wang[*]
Department of Physics and Astronomy
The Johns Hopkins University

## ABSTRACT

We study steady, radial gas outflows from galaxies in an effort to understand the way tenuous and hot gas is transported to large distances away from galaxies. In particular, we obtain solutions for outflow problems and study the outflow topology, the effects of the galaxy potential (and thus the mass), the size of outflow regions, the efficiency of radiative cooling, and the fate of the cooled gas. Under general power-law forms for the cooling function and the gravitational field of galaxies, we show that the outflow solutions are determined by the two-parameter initial conditions. In an analogy with stellar wind or accretion problems, we demonstrate that there exist no transonic flows, but either subsonic or supersonic flows are obtainable. Solutions of the supersonic outflows are studied in detail as they are most likely to carry gas to large distances away from galaxies. We find that, if gravity is weak, the outflow is characterized by the ratio of the radiative cooling time to the flow time, $t_c/t_f$. If initially $t_c/t_f \lesssim 1$ the gas cools as soon as it leaves the galaxy, whereas if $t_c/t_f \gtrsim 1$ the gas first cools adiabatically and then radiatively. However, if initially $t_c/t_f \gg 1$ radiation cannot become important; in this case, depending on whether the initial gas temperature is above the escape temperature, the outflow results in either a galactic wind or a hot corona. The importance of the galactic gravitational field is characterized by the fractional energy lost radiatively within the flow time in outflows with velocity equal to the circular velocity of the galaxy; if the fraction is small, gravity is relatively strong, and it stops the outflow before the gas has a chance to cool radiatively, resulting in a hot corona. In case the gas does cool radiatively, the cooled gas is most likely to form clouds via various instabilities. The clouds coast farther away from the galaxy because of the finite kinetic energy they inherit. Depending on the initial energies, the clouds can either leave the galaxy or fall back ballistically. We apply the calculations to galaxies in general. For reasonable ranges of the parameters, we find that the hot gas in dwarf galaxies can either flow out as galactic winds, or cool radiatively to form clouds. In the latter case, the clouds escape the galaxies. In contrast, massive galaxies like our own tend to confine the gas. The gas released into the halo can either cool radiatively or result in a galactic corona. We present the surface brightness in various x-ray energy bands for some representative cooling outflows from dwarf and normal galaxies, and we find that the extent of the resultant X-ray emission is generally much smaller than the region of the outflows. In particular, the mean temperature averaged over entire outflow regions is shown to be a factor of 2-6 smaller than the base temperature. We also estimate the mean surface brightness of the O VI emission lines, and the predicted surface brightness is within the reach of current UV experiments. We briefly discuss the implications of cold clouds at large distances from dwarf galaxies for recent observations of the QSO absorption line systems.

## 1. INTRODUCTION

Studies of gas outflows from galaxies are of considerable importance to understanding various galaxy phenomena. For example, the ejection of enriched gas dilutes the abundance of the interstellar gas in galaxies, thus affecting the subsequent rate of radiative cooling and star formation (e.g., Larson 1974; Wang & Silk 1993). Outflows can transport momentum and energy out of galaxies, therefore changing the dynamics of the galaxies (e.g., Dekel & Silk 1986). The heavy elements and kinetic energy injected from galaxies through outflows may have caused heating and enrichment of the intracluster medium and intergalactic medium in the early universe (e.g., White 1989).

Of more recent interest is the increasing realization that the gas found through QSO absorption lines such as the Mg II lines is associated with galaxies at extremely large distances. Indeed, attempts to identify spectroscopically the underlying galaxies at intermediate redshifts have been successful, indicating that the Mg II absorbing gas is at a typical distance of about 50 kpc from the galactic centers (e.g., Bergeron & Boisse 1991; Yanny *et al.* 1990; Bechtold & Ellingson 1992; Steidel 1993; Steidel & Dickinson 1995). More recently, studies of the Ly$\alpha$ clouds in relation to nearby galaxies suggest that they may be also associated with galaxies, possibly at distances even larger than the heavy element absorption line gas (Morris et al. 1993; Lanzetta et al. 1995). At high redshifts, observations of pairs of quasars show that the Ly$\alpha$ clouds have an average radius of about 100 kpc (Bechtold et al. 1994; Dinshaw et al. 1995).

---

[*] Also Space Telescope Science Institute

A possible explanation is that the absorbing gas results from gas outflows from the main body of the galaxies where most of stars and gas reside (Bregman 1981). Mathews & Baker (1971) first considered outward-flowing galactic winds from elliptical galaxies as a result of the gas ejection from stars. However, here we are mostly concerned with late-type (star-forming) galaxies, both because they are generally more numerous and because active star formation is inferred in the absorbing galaxies (e.g., Steidel 1993). In star-forming galaxies, the interstellar gas is shock-heated by stellar winds and more importantly, by supernova remnants to an extremely high temperature (e.g., Cox & Smith 1974). That the resulting hot gas may flow out of the galaxy was first proposed by Shapiro & Field (1976), and the possibility that the gas cools radiatively in the halo and subsequently falls back to the disk ballistically is called a "galactic fountain".

Subsequent hydrodynamic calculations of the galactic fountains have focused mostly on gas flows in galactic potentials similar to our Galaxy (Bregman 1980, 1981; Habe & Ikeuchi 1980; Houck & Bregman 1990; Li & Ikeuchi 1992; also see Cox 1981). However, simple arguments show that in galaxies like our own the heating rate required for the gas to reach a distance from the galactic center comparable to that observed in the metal line systems is implausibly large (Wang 1993). This is not surprising because the deep gravitational potential of galaxies requires extremely energetic gas to move out of the potential. This perhaps also explains the scarceness of the large X-ray halos detected for normal spiral galaxies like our own by ROSAT. However, for less massive (dwarf) galaxies one expects that gas can reach large distances because of the galaxies' shallow potential wells (e.g., Silk & Norman 1979; Fransson & Epstein 1982; Dekel & Silk 1986, Vader 1987).

Despite the increasing realization of the importance of gas outflows from galaxies, many questions concerning the physical nature of outflows remain unanswered. In particular, the dynamics of the outflows, such as the topology, the condition for radiative cooling to become important, the effects of gravitational field (thus the galaxy mass), and the extent of outflows, has not been studied systematically. As the first in a series to study gas outflows from galaxies, the present paper is aimed at obtaining exact solutions for the outflow problems and answering the above questions in a systematic way.

The temperature and volume filling factor of the hot phase in the interstellar medium (ISM) resulting from supernova heating depends on the supernova rate per unit volume and the density of the ISM, which are not precisely known even for our Galaxy. Thus we shall assume that the outflow originates from a fixed initial (base) radius at an initial temperature; the heating of the gas in the ISM is not considered here. Instead we consider ranges of the initial temperature and the initial radius as representing a variety of galaxies. If the porosity of the hot phase in the ISM is close to unity, the base radius is roughly the edge of the stellar distribution where the heating by supernovae decreases substantially. If the interstellar volume is dominated by cold clouds, the base radius should then be roughly the radius where the hot gas breaks out of the cold medium (such as in the chimney model; Norman & Ikeuchi 1989). We ignore thermal conduction, which may be unimportant in the large-scale dynamics of the upward flow since the heat flux conducted into the halo is usually smaller than the energy radiated (Houck & Bregman 1990). Any possible magnetic field (ignored here) will further impede thermal conduction.

We shall consider only steady outflows. Typically, the mass of the hot phase in the ISM is negligible compared with the total mass of the ISM, so a sufficient gas reservoir is generally available. The the steady flow calculations can be applied to galaxies where the duration of energy injection from supernovae into the ISM is sufficiently long, so the temperature at the base of the halo may be kept fixed during the flow, and a steady state can be achieved. For a velocity of 150 km/s (corresponding to the sound speed in a gas of $10^6$ K), the gas advances 15 kpc within $10^8$ years. A burst of star formation, even it is a $\delta$-function in time, will result in the gas heating with a duration of at least a few times $10^7$ years for the standard initial mass function because of the finite lifetime of the massive stars that eventually explode as supernovae (e.g., Leitherer & Heckman 1995). Thus the steady flow solutions can be applied to dwarf galaxies where the outflow region is typically less than a few kpc (see below), regardless of whether star formation is in a burst or in a continuous phase. They are also applicable to normal disk galaxies like our own, where star formation has been continuous and has varied little for the last few billion years (e.g., Sandage 1986; Wang & Silk 1994). Finally, we limit our calculations to radial (spherically symmetric) flows, which may be a reasonable approximation if the base of the flow is much smaller than the region of the outflow.

In §2 we first show that for power-law and logarithmic potentials where no additional scales are introduced, the outflow equations can be reduced to dimensionless form by introducing two natural scales in the problem (§2a). We then demonstrate that the outflow solutions cannot be transonic, in an analogy with stellar wind or accretion problems (§2b; Appendix), and that they can be obtained in terms of the two-parameter initial conditions. We then give examples of the supersonic solutions (§2c), and discuss the general properties of the outflows (§2d). In §3, we apply our calculations to dwarf and normal galaxies. Finally, we present a discussion and summary in §4.

## 2. COOLING OUTFLOWS

### 2a). Formulation of the Problem

The equations of mass, momentum, and energy conservation in a steady, spherically symmetric, flow are

$$\frac{1}{r^2} \frac{d\left(r^2 \rho v\right)}{dr} = 0 \tag{2.1}$$

$$v \frac{dv}{dr} = -\frac{1}{\rho} \frac{dp}{dr} - \frac{d\phi}{dr} \tag{2.2}$$

$$\frac{pv}{\gamma - 1} \frac{d}{dr} \ln \left(\frac{p}{\rho^\gamma}\right) = -C, \tag{2.3}$$

where $\rho$, $v$ and $p$ are, respectively, the density, velocity, and pressure of the gas, $\gamma$ (we take $\gamma = 5/3$ unless otherwise noted) is the adiabatic index, $\phi$ is gravitational potential (self-gravity of the gas is not included), and $C$ is the cooling rate per unit volume. We ignore heating from possible intergalactic UV background radiation since it is in general not important for the hot gas considered here. The local adiabatic sound speed is

$$c = \left(\frac{\gamma p}{\rho}\right)^{1/2}. \tag{2.4}$$

As $c^2$ is proportional to the gas temperature, $T_g$, the cooling rate per unit volume can then be written as

$$C = A\rho^2 c^{2q}. \tag{2.5}$$

For example, for a gas of near-cosmic abundance of heavy elements, cooling is dominated by bremsstrahlung at $T_g \gtrsim 5 \times 10^7 K$ so $q = 1/2$. At the lower temperatures, cooling is dominated by the line emissions, and $q \simeq -1/2$ to $-1$ (Dalgarno & McCray 1972; Raymond, Cox, & Smith 1976; Gaetz & Salpeter 1983; Sutherland & Dopita 1993). Using equation (2.1) to give

$$4\pi r^2 \rho v = \lambda = \text{const.}, \tag{2.1a}$$

where $\lambda$ is the (constant) mass-loss rate, and substituting $\rho$ in equations (2.2) and (2.3), we obtain

$$\frac{dv^2}{dr} = \frac{(1-\gamma)Ac^{2q}\lambda - 8\pi rc^2 v^2 + 4\pi r^2 v^2 (d\phi/dr)}{2\pi r^2 \left(c^2 - v^2\right)} \tag{2.6}$$

$$\frac{dc^2}{dr} = \frac{(1-\gamma)\left[(c^2 - \gamma v^2)Ac^{2q}\lambda - 8\pi r v^4 c^2 + 4\pi r^2 v^2 c^2 (d\phi/dr)\right]}{4\pi r^2 v^2 (c^2 - v^2)}. \tag{2.7}$$

It is thus clear from the above equations that the sonic point, where $v^2 = c^2$, is the critical point where the numerator vanishes, and the derivatives of $v$ and $c$ become infinite unless the numerators in equations (2.6)-(2.7) also vanish. If we use the Mach number, $M = v/c$, as the variable (e.g., Bondi 1952), then the critical point is at $M^2 = 1$. Notice that the above equations are invariant under the transformation $v \to -v$ (or $M \to -M$), so the equations describe both gas outflow and accretion.

We now consider some specific forms of galactic gravitational potential, which is related to the circular (or rotation) velocity of the galaxy, $v_{cir}$, by

$$r\frac{d\phi}{dr} \equiv v_{cir}^2. \tag{2.8}$$

For simplicity, we assume a rotation curve

$$v_{cir}^2 = \frac{B}{r^n}. \tag{2.9}$$

For example, for Keplerian orbit, $n = 1$ and $B = GM_G$, where $G$ is the gravitational constant, and $M_G$ is the total mass interior to the orbit. For galaxies dominated by dark halos with a flat rotation curve (isothermal spheres), we have $n = 0$ and $B = v_{cir}^2 = $ constant.

Given the gravitational potential in equation (2.9), there exist only two dimensional quantities in the present flow problem specified by equations (2.6)-(2.7): a characteristic length, $r_0$, and a characteristic velocity, $v_0$.[1] It is easy to show, from dimensional arguments using $A$, $B$, and $\lambda$, that these characteristic scales can be defined as the following:

$$r_0 = \left(\frac{A\lambda}{B^{2-q}}\right)^{1/(1-2n+nq)} \tag{2.10}$$

$$v_0 = \left(\frac{B}{A^n \lambda^n}\right)^{1/2(1-2n+nq)}. \tag{2.11}$$

The characteristic radius $r_0$ and velocity $v_0$ above contain information concerning the radiative cooling and the gravitational field of the galaxy. This is most obvious if we consider isothermal spheres ($n = 0$): equations (2.10) and (2.11) now read

$$r_0 = \frac{A v_{cir}^{2q} \lambda}{v_{cir}^4} \tag{2.10a}$$

$$v_0 = v_{cir}. \tag{2.11a}$$

Thus in this case, $r_0$ specifies the rate of the radiative cooling in a flow with velocity equal to the rotation velocity of the galaxy, and $v_0$ is the rotation velocity itself, determining the strength of the gravitational field. Note that the mass-flow rate enters here because it determines the density of the gas and therefore the cooling rate.

We can now define dimensionless variables in place of the radius and sound speed (or temperature):

$$x \equiv \frac{r}{r_0}; \tag{2.12}$$

$$w \equiv \frac{c}{v_0}. \tag{2.13}$$

The dimensionless radius, $x$, is a parameter measuring the strength of the gravitational field of the galaxy relative to the radiative cooling rate (just as $r_0$ is). In fact, if $n = 0$ we have

$$x = \frac{\rho v_{cir}^2}{4\pi (r/v_{cir}) C}. \tag{2.12a}$$

---

[1] The gas density can be found from mass conservation once solutions of (2.6)-(2.7) are obtained. The density scales as $\rho_0 = \lambda/(4\pi r_0^2 v_0)$, and the pressure scales as $p_0 = v_0 \lambda/(4\gamma \pi r_0^2)$.

That is, $1/x$ is the fraction of the energy lost to radiation during the flow time $r/v_{cir}$ for a flow with velocity equal to the rotation velocity of the galaxy. Therefore, if $x \gtrsim 1$ one expects that gravity becomes important in the flow. The dimensionless temperature, $w^2$, is a parameter characterizing the gas thermal energy relative to the potential energy. Thus for $w^2 \gg 1$ we expect that the gas escapes the galaxy.

Using the Mach number instead of the velocity, and substituting equations (2.12) and (2.13) into equations (2.6) and (2.7), we obtain the dimensionless equations for the Mach number and the dimensionless temperature:

$$\frac{dM^2}{dx} = \frac{(\gamma-1)\,x^n w^{2q}\left(1+\gamma M^2\right) + 8\pi x^{1+n} w^4 M^2 \left[2+(\gamma-1)M^2\right] - 4\pi(\gamma+1)\,x w^2 M^2}{4\pi x^{2+n} w^4 (M^2-1)} \qquad (2.14)$$

$$\frac{dw^2}{dx} = \frac{(\gamma-1)\left[-x^n w^{2q}\left(\gamma M^2-1\right) - 8\pi x^{1+n} w^4 M^4 + 4\pi x w^2 M^2\right]}{4\pi x^{2+n} w^2 M^2 (M^2-1)}. \qquad (2.15)$$

The first, second, and third terms in the numerators of equation (2.14) are due to radiative cooling, the gas pressure, and gravitation, respectively. Similarly, the first, second, and third terms in the numerator of equation (2.15) are contributions separately from radiative cooling, adiabatic expansion, and gravitation. The reduction to dimensionless equations above greatly simplifies our investigation of the flow solutions in relation to various galaxy properties (such as the mass-loss rate, the temperature at the base of the halo, the depth of the galactic potential, and so on). The complexity of taking into account varying galaxy parameters is reduced to a set of initial condition problems.

Two crucial timescales in the present problem are the flow time and the cooling time. We define the flow time as

$$t_f = \frac{r}{v}, \qquad (2.16)$$

and the cooling time as

$$t_c = \frac{p}{(\gamma-1)C}. \qquad (2.17)$$

The ratio of the cooling time to the flow time in terms of our dimensionless variables is:

$$\frac{t_c}{t_f} = \frac{4\pi}{\gamma(\gamma-1)} x w^{4-2q} M^2. \qquad (2.18)$$

In fact the above ratio is roughly the ratio of the contributions from adiabatic expansion and radiative cooling in supersonic flows described by equations (2.14) and (2.15) (i.e., the second and first terms in the numerators). If $t_c/t_f \gtrsim 1$, the flow is adiabatic, and if $t_c/t_f \lesssim 1$, radiative cooling is important, and the gas cools rapidly to low temperatures.

*2b). Supersonic Flow*

In the absence of radiative cooling (i.e., $q \to -\infty$), and for the point-mass potential (i.e., $n = 1$), equations (2.14) and (2.15) reduce to the classic Bondi problem (Bondi 1952); Bondi solved a spherical accretion problem for fixed adiabatic indices ranging from $\gamma = 1$ (isothermal) to $\gamma = 5/3$ (adiabatic). The corresponding wind solutions have been discussed by Parker (1958) in the context of the hydrodynamic model of the solar wind. From these classic studies, it is well understood that solutions for transonic flows exist only for $\gamma < 5/3$. There are no solutions proceeding smoothly from a subsonic to a supersonic regime owing to the stiff pressure response of an adiabatic gas; the sonic point progressively moves toward the origin as $\gamma \to 5/3$. Solutions do exist, however, for either subsonic or supersonic flows for $\gamma \geq 5/3$. The branch of the subsonic solutions turns out to require a finite confining pressure at infinity (Parker 1958).

For general potentials with arbitrary values of $n$, the topology of the solutions with constant $\gamma$ differs from the $n = 1$ case. In the Appendix, we present flow solutions for spherical accretion and wind problems at a fixed $\gamma$ (but without the cooling term included) in general potentials given by equation (2.9) with arbitrary values of $n$. Our main results there can be summarized as the following: (1) The range of $\gamma$ within which a transonic flow is possible is progressively reduced as $n$ decreases below unity. In particular, for isothermal spheres ($n = 0$), a transonic flow can occur only if $\gamma < 1$. Obviously, as the strength of the gravitational field increases toward small values of $n$, the gas finds it more and more difficult to overcome the sonic barrier unless heating is extremely effective. (2) For a given value of $n$, there exists a critical value of $\gamma$ (e.g., $\gamma = 5/3$, 1 for $n = 1$, 0, respectively), above which flow can only be either subsonic or supersonic. If the total energy $E > 0$, the gas can reach to infinity, with the Mach number increasing (decreasing) toward large radii for supersonic (subsonic) flows. However, in subsonic flows the density and temperature approach asymptotically constant values, implying a finite confining pressure at infinity, a generalization of the result found by Parker (1958) for the $n = 1$ case. (3) For supersonic or subsonic flows in the above case, if the total energy $E < 0$, there is a maximum radius beyond which no flow solutions exist. For $n \neq 0$, the flow starts either as subsonic and advances to the sonic point or as supersonic and abates to the sonic point, as the flow approaches the final radius where it is stopped by gravity. For isothermals spheres (n=0), however, there are two sonic points. For outflow problems, the gas has to start at the inner sonic point in both supersonic and subsonic flows, and it approaches the outer sonic point as the flow reaches the final radius.

With radiative cooling ($q =$ finite) (and $\gamma = 5/3$), pressure response to density variations is even steeper than in the adiabatic flow. This effectively increases the adiabatic index, and from the Appendix we see that there exists no transonic flow solution given the potential defined in equation (2.9). This indeed is borne out by our numerical analyses of equations (2.6)-(2.7). Solutions do exist, however, for either supersonic or subsonic flows. A potential problem with the subsonic flows is the outer boundary conditions mentioned above. Furthermore, because the velocities are relatively small in subsonic flows, the flows are close to hydrostatic equilibrium, as can be inferred directly from equation (2.2). Thus gravity is always important in subsonic flows. The entropy of the gas decreases toward large radii as a result of radiative cooling. Consequently, the subsonic flow is subject to convective and thermal instabilities (Balbus & Soker 1989). Hydrodynamic simulations of subsonic cooling outflows indeed bear this out (e.g., Houck & Bregman 1990). The supersonic flows, however, are driven by thermal pressure, and gravity is relatively unimportant for most of the flow. The gas becomes unstable only when it cools radiatively and gravity becomes important at the end of the flow. Various hydrodynamic simulations of supersonic cooling outflows show this is indeed the case (Bregman 1980; Habe & Ikeuchi 1980; Li & Ikeuchi 1992).

Aside from instabilities and possible outer boundary problems, subsonic flows have smaller velocities than their supersonic counterparts, and the resultant flow densities are higher for given mass-loss rate and initial gas temperature. Thus the radiative loss is more efficient and the resulting halo size is smaller than in the supersonic flows. As we are mostly interested in the largest extent of the galactic halos resulting from outflows, below we shall concentrate on the supersonic flows. Furthermore, we shall consider isothermals spheres ($n = 0$) for specific numerical examples since many galactic potentials can be well approximated by a flat rotation curve.

The initial Mach number of the flow is taken as unity at the base of the flow, a natural choice for the initial condition in supersonic (or subsonic) flows. Indeed, in a calculation of winds from starburst galaxies, Chevalier & Clegg (1985) obtained an analytical solution for the wind driven from a region of uniform mass and energy deposition; both gravitational field and radiative cooling were ignored but a heating term was added to the flow equations (cf. eqs.[2.1]-[2.3]). These conditions may be applicable to the region of the gas heating (the base); here the heating of the gas is more important than radiative cooling, a prerequisite for the existence of the hot phase in the ISM in the first place. Furthermore, the gravitational potential within the base should not vary largely because of the relatively small size of the heating region, and also because the potential is dominated by stellar populations rather than dark matter (so the variation of the potential is not as steep). Chevalier & Clegg found that a smooth transition from subsonic flow at the center to supersonic flow at large radius $r$ requires that $M = 1$ at $r = R$, where $R$ is the radius beyond which the heating ceases. The conditions imposed in the present work (i.e., inclusion of gravitational field and radiative cooling) are appropriate for the gas beyond the heating region, and thus our solutions can be considered as an extension of their inner ($r \leq R$) solutions; the transition is at $r = R$ with $M = 1$.

We thus see that solutions of equations (2.14) and (2.15) are determined by two initial conditions $x_i$ and $w_i^2$ at the base radius. These two parameters reflect the initial conditions of the gas, namely the relative importance of the gravitational potential to the radiative cooling, and of the internal thermal energy to the potential energy. Therefore, variations in $x_i$ and $w_i^2$ represent the different conditions of galaxies such as the region of the gas heating, the gas temperature and the mass of the galaxies.

*2c). Examples of Outflow Solutions*

As examples we show solutions for supersonic outflows with several values of the initial dimensionless temperature $w_i^2$ and the initial dimensionless radius $x_i$ in figures 1a-1c. The adopted values are ($x_i$, $w_i^2$)=($10^{-4}$, 25) (figure 1a), (0.6, 1) (figure 1b) and (10, 1) (figure 1c). Because equations (2.14) and (2.15) diverge at $M^2 = 1$, in practice we start the outward integration at $M^2 = 1 + \epsilon$, with $\epsilon \ll 1$ (typically $10^{-3}$) being sufficiently small so that backward integration to $M^2 = 1$ does not in practice change any initial conditions imposed. The cooling exponent is assumed to be $q = -0.6$, corresponding to a gas of nearly cosmic abundance at a temperature $10^5 \lesssim T_g \lesssim 4 \times 10^7$ K (Dalgarno & McCray 1972; Raymond, Cox, & Smith 1976; Gaetz & Salpeter 1983; Sutherland & Dopita 1993). Gravitational field of equations (2.8) and (2.9) with $n = 0$ is adopted.

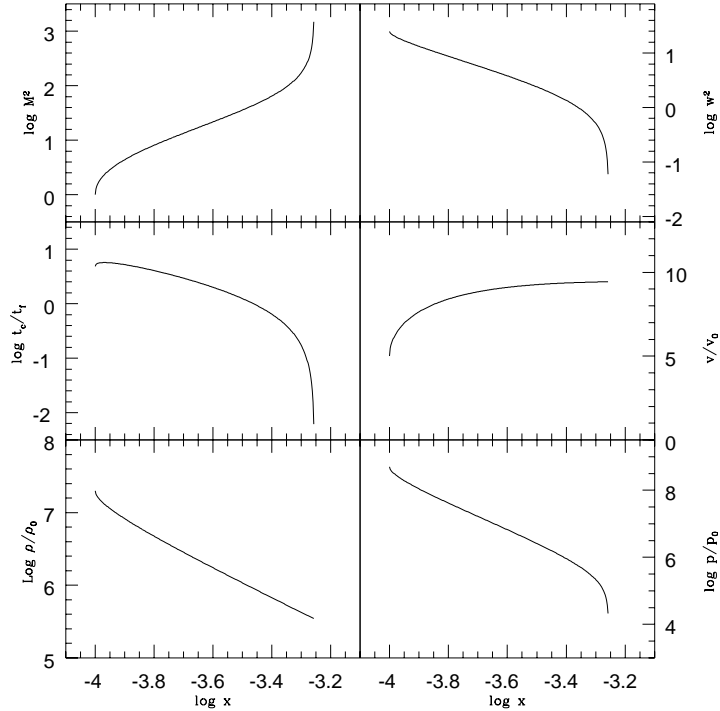

Fig. 1a

The two sets of ($x_i$, $w_i^2$) in figure 1a and 1b have a similar ratio of the cooling to the flow time, $t_c/t_f$, despite their large difference in the initial conditions. The values of ($x_i$, $w_i^2$) in figure 1c on the other hand lead to a much larger $t_c/t_f$ even though the initial dimensionless temperature $w_i^2$ is the same as in figure 1b; for given $w_i^2$ the relative importance of radiative cooling is reduced by increasing $x_i$ (e.g., by lowering the mass-loss rate). As can be seen in the figures, for the parameters considered here radiative cooling is initially unimportant in all the cases and the gas first cools almost entirely owing to adiabatic expansion. In figures 1a and 1b, as gas streams out toward large radii, radiative cooling eventually becomes important and the gas cools rapidly. In figure 1c, however, the time for radiative cooling is so long compared with the flow time that the flow remains adiabatic; here, before the gas has a chance to cool radiatively, the gas is pulled back by the strong gravitational field (characterized by $x_i > 1$), a situation already seen in the analytical solutions in the Appendix (cf. figure A3b). As the gas slows down and the density increases, the gas is in fact heated up through adiabatic compression. Because the flow time is much shorter than the cooling time in figure 1c, the radiative cooling term can be ignored. We then expect that the gas in the halo should reach a quasi-hydrostatic equilibrium over a few crossing times, resulting in a hot corona. Given enough time (many crossing times), the gas will cool gradually, with the cooled gas falling back to the galaxy. The amount of mass cooled is simply the lifetime times the cooling rate and should be much smaller than the total mass in the hot corona as long as the cooling time remains long.

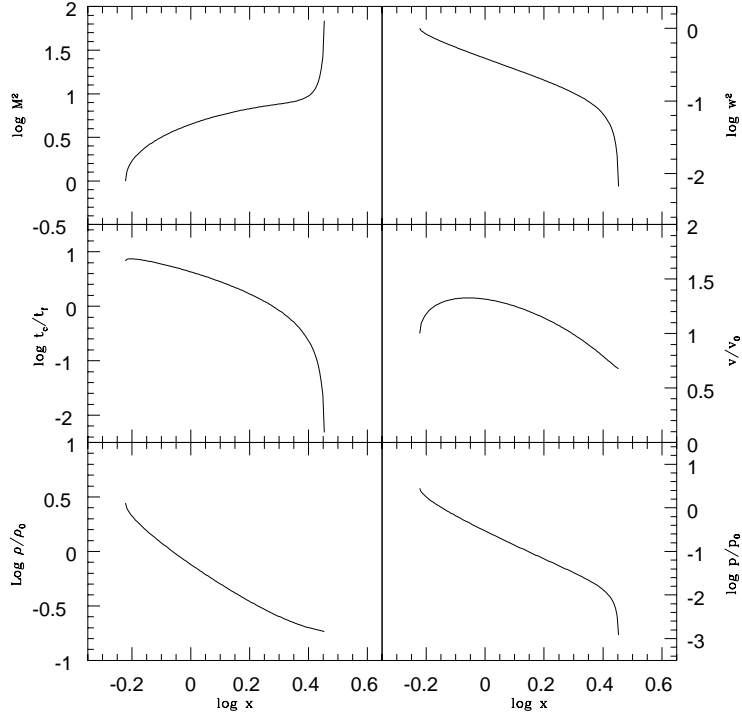

Fig. 1b

The primary driving force of the flow is thermal pressure; the gas is first accelerated by the gas pressure, which is much larger than gravity, and the thermal energy is converted to the kinetic energy. A unit mass of fluid carries with it an amount of internal energy $E_i = h + v^2/2$ (excluding the potential energy), where $h$ is the enthalpy. Thus initially the gas has an energy per unit mass of

$$E_i = \frac{\gamma + 1}{2(\gamma - 1)} c_i^2 = 2w_i^2 v_0^2, \tag{2.19}$$

where $c_i$ is the initial adiabatic sound speed. Part of the this energy has to be used to overcome gravitation, so the maximum velocity is less than $\sqrt{2E_i} = 2w_i v_0$. The gas cools as it expands, and the pressure drops. Eventually gravitation becomes dominant, and pulls the gas back. The actual peak of the velocity is reached roughly when the pressure force is equal to gravity; from equations (2.6) and (2.7) we find this occurs when $w^2 \simeq 1/2$ for a purely adiabatic flow. The energy loss to radiation can be estimated from the energy difference between the initial and final states of the flow. If $t_c/t_f$ is initially small, radiative cooling sets in relatively early, and radiative loss can be a significant part of the initial thermal and kinetic energy. Conversely, if $t_c/t_f$ is initially large, the time for the radiative loss is long, and when the gas finally cools radiatively, much of its internal energy has been used to overcome gravitation, so radiative loss is not a large fraction of the initial internal energy. For figures 1a and 1b, the energy lost to radiation is about 8% and 11% of the total internal energy in equation (2.19), respectively.

When the gas begins losing energy radiatively, it does so very rapidly owing to the increase of the cooling rate with decreasing temperature. The pressure of the gas drops rapidly, and now gravitation becomes dominant. At this point, because the entropy decreases outwardly as the gas cools radiatively, convective and thermal instabilities (Field 1965) ensue, and cloud formation follows. The criterion and some general properties of the convective and thermal instabilities in cooling flows have been worked out by Balbus & Soker (1989), but applications to the present problem require more careful and detailed analysis since supersonic motions may overrun the instabilities. We shall leave to a future paper the detailed stability analyses of the supersonic flow when radiative cooling becomes dominant. For the present purpose, however, it suffices to assume that, once radiative cooling sets in and the temperature of the gas drops precipitously (e.g., below $10^5$ K) (resulting from radiative cooling) cloud formation ensues, and it completes at cooling radius $x_c$ where the temperature drops to $\sim 10^4$. Note that radiative cooling is a prerequisite for cloud formation here; gas can still cool to a very low temperature via adiabatic expansion alone, but in this case the flow is stable and the resultant flow is a wind, similar to stellar winds. The onset of radiative cooling is signaled by a precipitous drop of $t_c/t_f$ below unity.

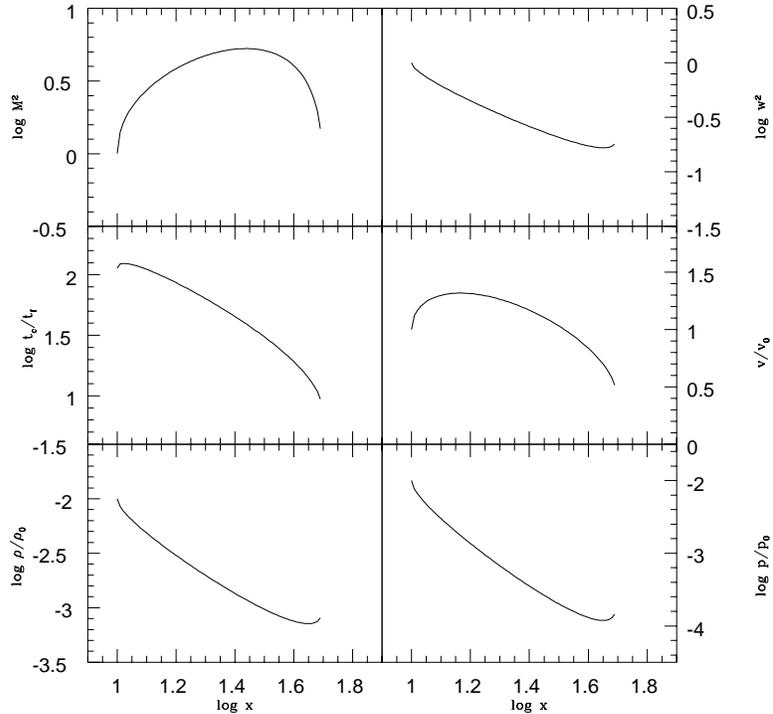

Fig. 1c

Figure 1. *a). Clockwise from top left, Mach number, dimensionless temperature, normalized velocity, normalized pressure, normalized density, and ratio of the cooling time to the flow time, as a function of the dimensionless radius. The initial conditions are $x_i = 10^{-4}$ and $w_i = 25$. b). Same as in a), but with $x_i = 0.6$ and $w_i = 1$. c). Same as in a), but with $x_i = 10$ and $w_i = 1$.*

One may associate the radius at which the temperature drops precipitously because of radiative loss with the location of cloud formation. Once clouds are formed, they drop out of the flow and move ballistically. The newly formed clouds retain the kinetic energy of the gas at the point of formation. Because the gas loses energy to radiation within a time much shorter than the flow time, it has moved little during the period of the rapid radiative cooling. Consequently, the velocity, which is mainly affected by the potential and thus the relative position of the gas in the halo, remains almost unchanged at the point of cloud formation. We may adopt the final velocity of the gas (where radiative cooling becomes dominant) as the initial velocity of the newly formed clouds. Whether these cooled clouds will eventually escape the galaxy or fall back under the influence of the galactic potential depends on the initial cloud velocity relative to the escape velocity, which in turn depends on the galaxy mass distribution that gives rise to the rotation curve.

If the mass distribution (and thus the flat rotation curve) is truncated at radius $r_{max}$, the resultant gravitational potential is given by equation (A2). In terms of the dimensionless radius, the escape velocity

is then
$$v_{esc}(x) = \sqrt{2}v_0 \left[1 + \ln\left(\frac{x_{max}}{x}\right)\right]^{1/2}. \tag{2.20}$$

For example, in figures 1a and 1b the radii at which the temperature of the gas drops precipitously because of the radiative loss are $x_c = 5.5 \times 10^{-4}$ and $x_c = 2.8$, and clouds formed at these radii have velocities 9.4 and 0.68 times $v_0$, respectively. If $x_{max} = 50x_i$ (or equivalently $r_{max} = 50r_i$), the velocities correspond to about 370% and 26% of the escape velocity, respectively, so the clouds will be confined in the halo in figure 1b but will leave the galaxy in figure 1a. If the gas cannot escape from the galaxy, the final radius to which the clouds can coast before being pulled back by gravitation can be estimated from

$$x_f = x_c \exp\left[\frac{1}{2}\left(\frac{v_c}{v_0}\right)^2\right], \tag{2.21}$$

if $x_f \leq x_{max}$, or

$$x_f = \frac{2v_0^2}{v_{esc}^2(x_c) - v_c^2}x_{max}, \tag{2.22}$$

if $x_f \geq x_{max}$, where $v_c$ is the initial velocity of the clouds at cooling radius $x_c$. In figure 1b, we have $x_f = 3.5$, or about 30% more distant than $x_c$.

Also plotted in figures 1a-1c are the density, in units of $\rho_0 = \lambda/(4\pi r_0^2 v_0)$, and the pressure, in units of $p_0 = v_0\lambda/(4\pi r_0^2)$. The density, after a sharp initial drop as a result of the velocity increase, decreases very roughly as $x^{-2}$ during adiabatic expansion since the velocity then does not vary much. In figure 1c, where the gas cannot cool radiatively, the density at the end of the outflow has a small jump as the flow is stopped by gravitation. A similar jump is also observed in the pressure. In contrast, in figures 1a and 1b, the pressure decreases sharply at the final stage of the outflow because the gas temperature drops precipitously as a result of radiative cooling.

So far we have assumed that the radiative cooling occurs within the edge of the mass distribution of the galaxy, i.e., $x_c \leq x_{max}$. If $x_c > x_{max}$, the gravitational force suffers a jump during the outflow; inside $x_{max}$ gravitational force is given by the logarithmic potential in equation (A2), and outside $r_{max}$ given by the potential in equation (A1) with $n = 1$. In this case, one can integrate flow equations (2.14) and (2.15) up to $x_{max}$, as done above. Switching to the $n = 1$ potential at $r_{max}$, and adopting as the initial conditions the values of the first part solutions at $x_{max}$, we can continue integration of the same equations from $r_{max}$ outward. (A careful matching at $x_{max}$ in terms of the dimensionless variables $x$ and $w$ is necessary because the length and velocity scales in eq.[2.10] and [2.11] are different for different values of $n$.) The effects of mass distribution of galaxies can then be investigated by varying $x_{max}$. The calculations are straightforward, and for clarity we only summarize the main results here.

If we denote $x_c$ by the cooling radius obtained by assuming $x_c < x_{max}$, we find that $x_c$ for a given $w_i$ is hardly affected by decreasing $x_{max}$ below $x_c$. This is expected because whether the gas can radiatively cool depends on the ratio of the radiative cooling time and the flow time, equation (2.18). The former is primarily determined by the initial conditions, and the latter is not much affected by gravitation during the adiabatic expansion in a supersonic flow. Thus for example, in figures 1a and 1b, $x_c$ is increased only by 0.4% and 27%, respectively, *even if we reduce $x_{max}$ to $x_i$, i.e., no massive dark halos for galaxies.* Changing $r_{max}$, however, does affect the fate of the cooled gas since it determines the escape velocity and, hence, whether the cooled gas will stay in or leave the galaxy. Similarly, it also influences the nature of the outflow if radiative cooling cannot occur, because $r_{max}$ in effect determines the total energy (including the potential energy) of the gas in the flow. As seen in the analytical solutions for outflows without radiative cooling in the Appendix, the topology of the flows is specified by the sign of the total energy. For example, in figure 1c, where radiative cooling is not important, if $x_{max}$ is reduced below about $x_{max} \simeq 27$, the outflow topology is changed; in this case the total energy of the flow becomes positive, and the gas leaves the galaxy. The Mach number now behaves as that in figure A2 in the Appendix. In the following, we shall be mainly concerned with the gas that can cool radiatively; the outflow solutions without radiative cooling can be readily found in the Appendix. We ignore the slight change of the cooling radius $x_c$ caused by varying $r_{max}$. However, we take into account the effect of finite $r_{max}$ on how far the cooled clouds can coast in the halo using equations (2.21)-(2.22).

*2d.) General Properties of the Outflow*

From the previous examples, we see that once radiative cooling sets in, the gas temperature drops precipitously almost at a fixed radius. This cooling radius $x_c$ for given initial dimensionless temperature $w_i^2$ is shown in figure 2 for several values of the dimensionless radius $x_i = 10^{-5}$, $10^{-4}$, $10^{-3}$, $10^{-2}$, $10^{-1}$, and 1 (solid curves). (Numerically, $x_c$ is determined when $t_c/t_f$ in equation [2.18] drops below 0.1.) In the limit $w_i^2 \to 0$, the gas is trapped in the host galaxy, and $x_c$ is equal to $x_i$. As $w_i^2$ increases, $x_c$ increases, and asymptotically approaches $\propto w_i^{4.59}$ for large $w_i^2$. The transition from the gradual to the asymptotic rise of $x_c$ occurs at initial $t_c/t_f \sim 1$; below values of $w_i^2$ that correspond to $t_c/t_f \sim 1$ at $x_i$, the gas cools as soon as it leaves the base of the galaxy, whereas above those values the gas streams out adiabatically first before cooling radiatively. The asymptotic behavior at large $w_i^2$ can be understood as the following. Before the onset of radiative cooling the outflow is approximately adiabatic. From equations (2.14) and (2.15), we find approximately

$$M^2 \sim (x/x_i)^{4/3}; \tag{2.23}$$

$$w^2 \sim w_i^2 (x/x_i)^{-4/3}, \tag{2.24}$$

if radiative cooling and gravity (the first and third terms in the numerators in eq. [2.14] and [2.15]) can be ignored. Radiative cooling becomes dominant when $t_c/t_f \lesssim 1$. So substituting equations (2.23) and (2.24) into equation (2.18), we find that the cooling radius asymptotically approaches

$$x_c \propto x_i^{\frac{4(1-q)}{1-4q}} w_i^{\frac{6(2-q)}{1-4q}} = x_i^{1.88} w_i^{4.59}, \tag{2.25}$$

where the last equality follows from $q = -0.6$.

As seen in figures 1a and 1b, even if the radiative cooling time is longer than the flow time initially, adiabatic expansion brings the gas temperature down, and with the assumed cooling function increasing with decreasing temperatures, radiative cooling takes over eventually. The gas then cools rapidly unless it is stopped prematurely by gravitation (e.g., figure 1c). In practice, however, if the temperature drop through adiabatic expansion is too large, our assumed cooling function may not be applicable. For example, if the gas cools adiabatically below $10^5$ K before radiative cooling can set in, the rate of radiative cooling for such a gas is smaller than the simple extrapolation from the higher temperatures given by equation (2.5). In fact, $q$ becomes positive ($\sim 0.5$ between $10^5$ and $10^4$ K) so the radiative cooling time is increased as the temperature decreases. As a result, the gas cannot cool radiatively. In this case, if the initial temperature of the gas is above the escape temperature, then the outflow becomes a wind, a situation similar to stellar winds. On the other hand, if the initial temperature is below the escape temperature, the gas remains in the galaxy, resulting in a hot corona.

The cooling function in equation (2.5) is roughly applicable within $10^5 - 4 \times 10^7$ K. The initial gas temperature of interest is above a few times $10^6$ K based on X-ray observations (see §3). For example, if the initial temperature is $10^7$ K, and radiative cooling cannot dominate before the adiabatic cooling brings the gas temperature down by a factor of 100, the gas will never cool radiatively. From equation (2.24), we find that the corresponding upper limits on the calculated cooling radius is about a factor of 32 times the initial radius; the cooling radius obtained beyond this limit is fortuitous because the adopted cooling function cannot apply here. In figure 2, we plot the values of $w_i^2$ above which radiative cooling cannot become dominant before the gas temperature drops by a factor of 100 through adiabatic expansion (shaded area above the dash-dotted line). If the initial temperature is higher (lower) than $10^7$ K, the corresponding $w_i^2$ for given $x_i$ is larger (smaller) than shown in figure 2. If the initial temperature is $\alpha \times 10^7$ K, the corresponding limit on $w_i^2$ changes by a factor of $\alpha^{(1-4q)/4(2-q)} = \alpha^{0.33}$ compared with the $10^7$ K case from equations (2.24) and (2.25). The above upper limits on $w_i^2$ can also be understood in terms of $t_c/t_f$; the limits correspond to $t_c/t_f = 42$ at the base of the outflow. That is, if the initial $t_c/t_f$ is larger, the radiative cooling time is so long that radiative cooling cannot become important.

The region of initial conditions ($x_i$, $w_i^2$) that results in wind is denoted by WINDS in figure 2; it is bounded below by the requirement that $w_i^2$ be above the escape temperature, so the gas can escape the galaxy. In the absence of radiative cooling the initial temperature necessary for the gas to escape is

$$T_{esc} = \frac{(\gamma-1)}{\gamma(\gamma+1)} \frac{\mu m_H}{k} v_{esc}^2(x_i), \tag{2.26}$$

where $\mu$ is the mean molecular weight, $m_H$ is the mass of hydrogen, and $\phi(x_i)$ is the potential at $x_i$. If we assume $x_{max} = 50 x_i$, the wind region should be above $w_i^2 = w_{esc}^2 = 2.456$ (dashed horizontal line) in figure 2. Below the wind region the gas cannot leave the galaxy, resulting in a hot corona.

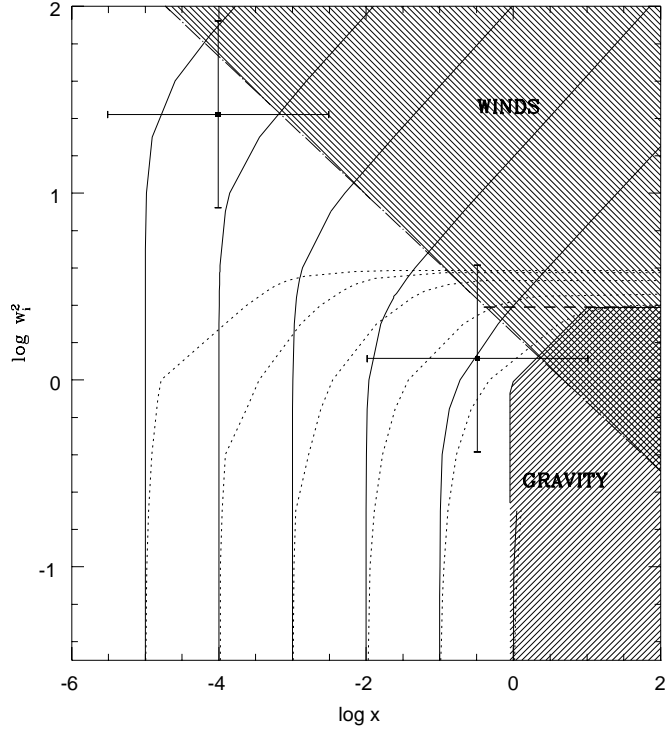

Fig. 2

Figure 2. *Initial dimensionless temperatures versus the dimensionless radius. The solid curves represent the cooling radius, $x_c$, for given $w_i^2$, calculated for several values of the initial radius $x_i$, from left to right: $x_i = 10^{-5}$, $10^{-3}$, $10^{-3}$, $10^{-2}$, $10^{-1}$, and 1. The dotted lines denote the final radius $x_f$ which clouds (formed at $x_c$) can reach for given $w_i^2$. The initial radii are the same as for the solid curves. Above certain values of $w_i^2$, $x_f$ becomes infinite, indicating that the clouds escape the galaxies. For all the cases, the mass distribution of the galaxies is truncated at $x_{max} = 50x_i$. The dash-dotted line shows the boundary of $w_i^2$ as a function of $x_i$ above which the initial conditions are such that radiative cooling cannot be important throughout the flow (shaded area). The shaded area is further divided into two regions; above the escape temperature $w_i^2 = 2.456$ (dashed horizontal line), the outflow results in galactic winds (denoted WINDS), and below it the outflows lead to coronae. The other shaded area in the lower right, denoted GRAVITY, represents the region of initial condition $(x_i, w_i^2)$ where the gravitational potential of the galaxies dominates and the gas is stopped by gravity before it has a chance to cool radiatively; in this case, the outflows also result in galactic coronae. The two crosses show the range of initial conditions $(x_i, w_i^2)$ for dwarf galaxies with rotation velocity $v_{cir} = 50$ km/s (left cross) and normal galaxies with rotation velocity $v_{cir} = 225$ km/s (right cross). The vertical extension of the crosses represents the initial temperature range considered, $T_i = 10^{6.5 \pm 0.5}$, and the horizontal stretch of the crosses represents the range $r_{i,kpc}/\lambda_1 = 10^{\pm 1.5}$ for both dwarf and normal galaxies.*

For $x_i \gtrsim 1$, gravitational force is strong relative to radiative cooling as discussed in §2a, so the gas will be pulled back to the galaxy before it has a chance to cool radiatively (cf., figure 1c). In figure 2, we show the region of $(x_i, w_i^2)$ where the gas is confined to the halo by gravitation and radiative cooling is not important during the flow time. The region is bounded below by $t_c/t_f \sim 1$ since obviously radiative cooling must not be too strong for gas to rise at all. It is also bounded above by the escape temperature, as the gas can leave the galaxy in this case. The escape velocity is taken to be that given by $x_{max} = 50x_i$.

We have seen in §2c that in case the gas cools radiatively and the temperature drops precipitously at the end, the cooled gas can coast farther beyond the cooling radius $x_c$ because the newly formed clouds inherit the finite kinetic energy from the hot gas. In figure 2 we show for given $w_i^2$ the maximum radius, $x_f$, which the cold clouds can eventually reach, based on equations (2.21) and (2.22) (dotted curves). Calculations are made for several initial radii $x_i = 10^{-5}$, $10^{-4}$, $10^{-3}$, $10^{-2}$, $10^{-1}$, and 1. For definiteness, we take $x_{max} = 50x_i$; in general, if $x_f \leq x_{max}$, $x_f$ is independent of $x_{max}$, and if $x_f > x_{max}$, $x_f$ is proportional to $x_{max}$, (eqs. [2.21] and [2.22]). An indication of the relative importance of radiative loss can be found from the minimum $w_i^2$ that allows the cooled gas to escape the galaxy. For example, in figure 2 we find that for $x_i = 10^{-5}$ the initial temperature required for escape is $w_i^2 \gtrsim 3.9$. The dimensionless escape temperature from equation (2.26) is $w_{esc}^2 = 2.456$ for $x_{max} = 50x_i$. Thus in case $w_i = 3.9$, about 60% of the initial internal energy is lost to radiation. As $x_i$ increases, the relative magnitude of radiative cooling diminishes. Much of the energy is consumed in overcoming gravitation, so radiative loss is only a small fraction of the initial internal energy. For instance, from figure 2 we note that for $x_i = 0.1$ the required $w_i^2$ for the escape is $\gtrsim 2.55$. In this case, we infer that only about 4% of the initial internal energy is lost to radiation.

## 3. APPLICATIONS

In this section, we switch to dimensional variables and make attempts to connect with various observations. More specifically, we first examine the difference in characteristics of the cooling outflows between dwarf and normal galaxies. We then calculate the X-ray surface brightness, the mean gas temperature, and the mean surface brightness of the O VI emission lines for some representative outflow solutions.

### 3a). Dwarf and Normal Galaxies

The cooling function for a gas of nearly cosmic abundance at $10^5 \lesssim T_g \lesssim 4 \times 10^7$ K may be taken approximately as (e.g., Sutherland & Dopita 1993)

$$q = -0.6$$
$$A = 3.4 \times 10^{34} \text{cgs}. \tag{3.1}$$

For gas with lower abundance, the above $q$ may still be roughly applied, but $A$ is smaller. For example, $A$ is a factor of about 3 (5) smaller for a gas of 1/10 (1/100) the cosmic abundance (e.g., Bohringer & Hensler 1989; Sutherland & Dopita 1993). Non equilibrium of collisional ionization may also reduce the cooling rate at a temperature below about $4 \times 10^5$ K. The reduction may be up to a factor of 4 at $10^5$ K (e.g., Shapiro & Moore 1976; Schmutzler & Tscharnuter 1993; Sutherland & Dopita 1993). If a cooling rate other than equation (3.1) is to be used, the corresponding results are modified according to figure 2; $x$ scales with $A\lambda$ (so reducing $A$ is equivalent to increasing $\lambda$).

Assuming a flat rotation curve ($n = 0$), from equations (2.10a) and (2.11a), we obtain

$$r_0 = 278 \frac{\lambda_1}{v_{100}^{5.2}} \text{kpc}, \tag{3.2}$$

where $\lambda_1 = \lambda/(M_\odot/\text{yr})$, and $v_{100} = v_0/(100 \text{km/s})$. Thus the dimensionless radius is

$$x = 3.6 \times 10^{-3} \frac{v_{100}^{5.2} r_{kpc}}{\lambda_1}, \tag{3.3}$$

and the dimensionless temperature is

$$w^2 = 2.2 \frac{T_6}{v_{100}^2}, \tag{3.4}$$

where $r_{kpc}$ is the radius in kpc, and $T_6 = T_g/10^6$ K. Note that equation (3.3) only depends on ratio $r_{kpc}/\lambda_1$. Now apply equations (3.3)-(3.4) to initial conditions $x_i$ and $w_i^2$ where $r = r_i$ and $T_g = T_i$. The initial gas temperature is primarily determined by the heating of the ISM by supernova remnants in the galaxy, which depends on the internal star formation rate per unit volume (e.g., McKee 1990). Observationally, the star formation rate is found to correlate more with morphological types rather than the mass of galaxies (e.g., Sandage 1986), so one may expect that the initial temperature is not a strong function of the galaxy mass. However, $x_i$ is a strong function of the circular velocity, and, thus of the mass of the galaxy. As a result, for a given mass-loss rate, massive galaxies in general have large $x_i$ and small $w_i$, and dwarf galaxies in general cluster around small $x_i$ and large $w_i$.

As examples, in figure 2 we show the inferred $(x_i, w_i^2)$ for dwarf galaxies with a circular velocity of 50 km/s if $T_{i,6} = 3$ and $r_{i,kpc}/\lambda_1 = 1$ (e.g., $r_{i,kpc} = 0.3$ and $\lambda_1 = 0.3$), where $T_{i,6} = T_i/10^6$K and $r_{i,kpc} = r_i/\text{kpc}$.

The cross on the left in figure 2 represents the uncertainties of the various parameters; the vertical extension of the cross results from the adopted temperature range $T_{i,6} = 10^{0.5 \pm 0.5}$, and the horizontal stretch reflects the adopted of $r_{i,kpc}/\lambda_1 = 10^{\pm 1.5}$ (e.g., $\lambda_1 = 0.01 - 10$ for fixed $r_{i,kpc} = 0.3$). Similarly, we also present the initial conditions expected for normal galaxies like our own with circular velocity $v_{cir} = 225$ km/s if $T_{i,6} = 3$, $r_{i,kpc}/\lambda_1 = 1$ (e.g., $r_{i,kpc} = 3$ and $\lambda_1 = 3$). A similar range of the parameters is taken: $T_{i,6} = 10^{0.5 \pm 0.5}$, and $r_{i,kpc}/\lambda_1 = 10^{\pm 1.5}$ (e.g., $\lambda_1 = 0.1 - 100$ for fixed $r_{i,kpc} = 3$) (right cross).

A comparison with the previous calculations of the cooling radius $x_c$ and the final radius $x_f$ for the cooled clouds in figure 2 shows that dwarf galaxies in general should have halos of hot gas many times their initial radii. Depending on the initial temperature and density, the gas can either cool or result in galactic winds. In either case, the gas is most likely to leave the galaxy because the temperature is usually above the escape temperature. In contrast, for massive galaxies like our own, unless the temperature is extremely high, the gas is in general confined within the galaxy. If the gas density is high enough (large $\lambda$ and thus small $x_i$), radiative cooling becomes important at the end and the gas cools in the halo. But typically the cooled gas is unable to travel much farther from the cooling radius because of the strong potential. If the gas temperature is below the escape temperature, and the density is low, the gas will be pulled back to the galaxy before it has a chance to cool radiatively. In this case, the flow will result in a galactic corona.

*3b). Surface Brightness and Temperature Structure*

We show in figure 3 some examples of the temperature structures and density profiles in the cooling outflows from dwarf galaxies with $v_{cir} = 50$ km/s, and from normal galaxies with $v_{cir} = 225$ km/s. We take $r_{i,kpc}$ roughly as the region of active star formation in galaxies. We consider a modest outflow with $T_i = 3 \times 10^6$ K, and a strong outflow with $T_i = 2 \times 10^7$ K. The input parameters and the general results of the outflow calculations are summarized in Table 1.

We also present the surface brightness in 0.1-2.2 keV (roughly the ROSAT energy range) and in 1.6-8.3 keV bands, and the total surface brightness. The 1.6-8.3 keV hard band may be compared with the X-ray observations by GINGA (2-20 keV), BBXRT(0.5-10 keV) and ASCA (0.5-10 keV). The surface brightness is calculated through the integral $F(r) = \int_r^\infty z\rho^2 \epsilon(T_g)/(2\pi\sqrt{z^2 - r^2})dz$, where the emissivity of the gas, $\epsilon(T_g)$, is taken from Suchkov et al. (1994).

If the outflows are dominated by adiabatic expansion, the temperature is proportional to $r^{-4/3}$, and the density is roughly $\rho \propto r^{-2}$, so the surface brightness is $\sim r\rho^2\epsilon \propto r^{3-4q/3} = r^{-2.2}$. For the modest outflows considered above, the initial cooling time is slightly larger than the flow time ($t_c/t_f = 5$), so radiative cooling soon becomes important as the flow streams out. As a result, the total surface brightness is much shallower than $r^{-2.2}$ because the temperature drops more rapidly than $r^{-4/3}$. The surface brightness in the ROSAT band, however, drops much more sharply than the total surface brightness because of the temperature decrease. The gas is also too cool to be easily seen in the hard energy band. In contrast, for the strong outflow, the initial cooling time is relatively large compared with the flow time ($t_c/t_f = 20$), so the flow first expands adiabatically, and the total surface brightness decreases roughly as $r^{-2.2}$. At the end of the flow, however, the cooling time becomes much shorter than the flow time, and the temperature drops rapidly as a result of radiative cooling. Consequently, the total surface brightness decreases less steeply than in the adiabatic phase before the final precipitous decrease.

The cooled clouds formed in both the modest and strong outflows from dwarf galaxies have large initial velocities well beyond the escape velocity for any reasonable mass distribution, so they leave the galaxies. For the normal galaxies, the escape velocities are 461, 514, 578, and 613 km/s if the mass distribution of the galaxies is truncated at 30, 50, 100, and 150 kpc, respectively. From Table 1, we thus see that the cooled gas is most likely to stay in the galaxies in the modest outflows, and it leaves galaxies in the strong outflows.

The calculations of the modest outflow from dwarf galaxies may be compared with the ROSAT observations of NGC 1569 ($v_{cir} = 50$ km/s) (Heckman et al. 1995). The extent of the observed X-ray halo emission is about several hundreds pc in radius with a luminosity in the ROSAT band of about $4 \times 10^{38}$ erg/s. The H$\alpha$ emission in NGC 1569 has an extent of about 200-300 pc in the central region (Walker 1991). The calculations of the modest flows from normal galaxies may be compared with the ROSAT observations of NGC 891 (Bregman & Pildis 1994). The full extent of the observed X-ray halo emission is about 7 kpc in radius with a luminosity in the ROSAT band of about $4 \times 10^{39}$ erg/s. The scale height of the extended H$\alpha$ emission in the central region of NGC 891 is about 3-4 kpc (Rand, Kulkarni, & Hester 1990; Dettmar 1990). Extended X-ray emissions from a few other galaxies have also been detected with ROSAT, including M82 (Bregman, Schulman, & Tomisaka 1995) and NGC 4631 (Wang et al. 1995). We shall present more detailed comparisons with observations elsewhere.

To compare with the average temperature of the entire halo derived from X-ray observations, we calculate the mean temperature averaged over the outflow region and weighted by the emission: $<T> = \int 4\pi r^2 C T_g dr / (\int 4\pi r^2 C dr)$. Figure 4 shows $<T>$ as a function of the initial temperature $T_i$ for the four examples shown in figure 3: Not surprisingly, the mean temperature is systematically lower than the initial gas temperature (by a factor of 2-6) because of the cooling of the gas away from the galaxies. Correctly taking this into account is important in understanding various physical conditions (such as the heating of the gas) in the host galaxies.

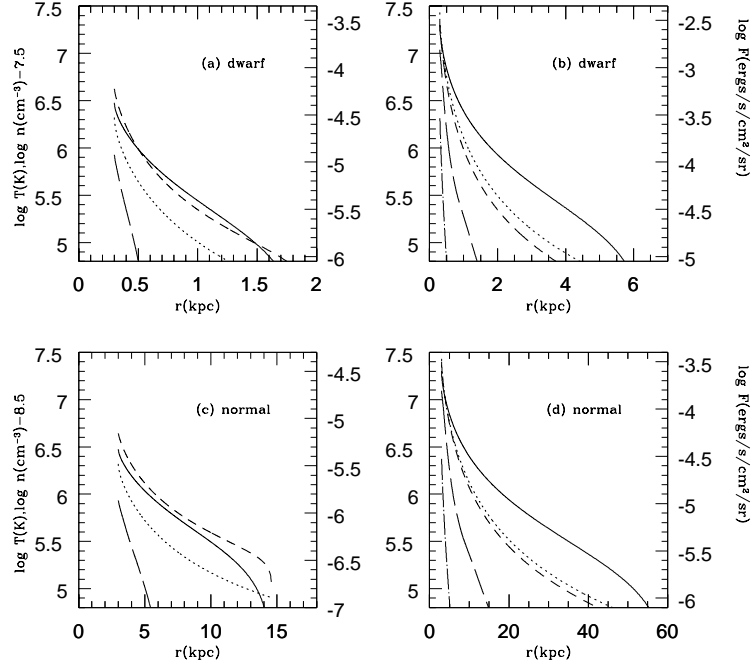

Fig. 3

Figure 3 *Gas temperature (solid curves), gas density (dotted), total surface brightness (dashed), surface brightness in 0.1-2.2 keV (long-dashed), and surface brightness in 1.6-8.3 keV (dot-dashed) as a function of the radius for some representative cooling outflows. The level of the surface brightness is labeled on the right axes. a). Results of the outflow from dwarf galaxies ($v_{cir} = 50$ km/s), with initial temperature $T_6 = 3$ and mass-loss rate $\lambda_1 = 0.3$. b). Same as a), but with $T_i = 20$ and $\lambda_1 = 10$. c). Results of the outflow from normal galaxies ($v_{cir} = 225$ km/s), with $T_6 = 3$ and $\lambda_1 = 3$. d). Same as c), but with $T_i = 20$ and $\lambda_1 = 100$. Other model parameters can be found in Table 1. In a) and c), the fluxes in 0.1-2.2 keV are below the flux ranges shown.*

The cooling gas can also be detected via emission lines, such as the O VI doublets ($\lambda 1034$ Å). For examples, for a gas cooling from $T_6 = 3$ ($T_6 = 20$) down to $T_6 = 0.1$, the energy emitted in the O VI emission lines is about 3% (0.5%) of the total energy lost to radiation (Voit, Donahue, & Slavin 1994). Table 1 gives the mean surface brightness in O VI, calculated as the surface brightness averaged over the flow region inside $r_c$; the actual surface brightness in the inner regions should be much higher than the average. The predicted mean surface brightness may be within the reach of current UV experiments. For example, the detectable O VI line flux (3 $\sigma$) of the spectrometer recently flown on a rocket is $2 \times 10^{-7}$ erg/s/cm$^2$/sr (Rasmussen & Martin 1992). The Hopkins Ultraviolet Telescope (HUT) flown aboard the Space Shuttle on the Astro-1 mission also has a similar sensitivity for a typical exposure time ($\sim 2000$ s) (Davidsen et al. 1992). Since HUT on Astro-2 mission scheduled next year is expected to improve the sensitivity by a factor of few (W. P. Blair 1994, private communication), the outflows, especially from dwarf galaxies, should be detectable by HUT.

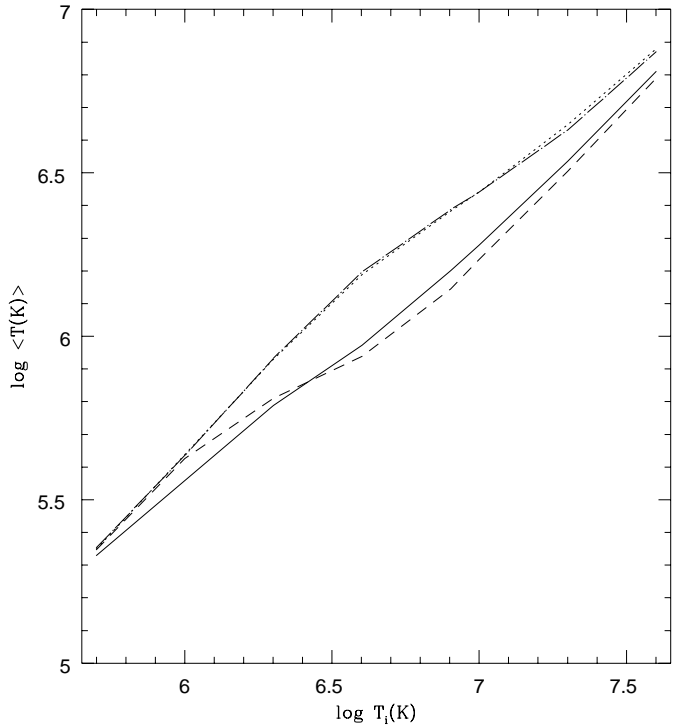

Fig. 4

Figure 4. *Mean temperature averaged over the entire flow region as a function of the initial temperature (eq. [3.7]), for the outflows shown in figures 3a (solid curve), 3b (dotted), 3c (dashed) and 3d (dot-dashed).*

## 4. DISCUSSION AND CONCLUSION

We study the outflows of hot gas from galaxies. In particular, we investigate the general properties of gas outflows. Emphasis is made on understanding the outflow topology, the extent of the outflow, the efficiency of radiative cooling, and the fate of the cooled gas. As a first step, we consider problems of steady, radial outflows with initial conditions specified at the base where the flow originates. Gravitational potentials in galaxies are assumed to have either power-law or logarithmic dependence on the radius. We demonstrate in analogy with stellar wind problems that without efficient heating the outflow cannot overcome the sonic barrier, so the outflow remains either subsonic or supersonic (Appendix). While subsonic flows may be subject to various convective and thermal instabilities and may not travel much farther from the base because of the relatively small velocities, the branch of supersonic solutions provides the most promising route for gas to reach large distances away from the galaxies.

We approximate the cooling function by a power-law of the gas temperature. Along with the assumed forms of the galactic potential, there exist two natural scales in the problem (eq. [2.10] and [2.11]), and the flow equations can be reduced to dimensionless form (eq. [2.14] and [2.15]). The outflow problems then can be formulated in terms of solving ordinary differential equations with two-parameter initial conditions: the dimensionless initial radius and the dimensionless initial temperature. This greatly reduces the parameter space one would need to explore. The dimensionless radius $x$ (eq. [2.12]) contains the information concerning the importance of the gravitational potential relative to radiative cooling, and the dimensionless temperature $w^2$ (eq. [2.13]) measures the thermal energy of the gas relative to the potential energy.

We obtain supersonic solutions for the various initial conditions representing a variety of galaxies. A few examples of those solutions are shown in figure 1. The characteristics of the outflows are determined by the ratio of the cooling time to the flow time, $t_c/t_f$ (eq. [2.18]), and by the initial dimensionless radius $x_i$. If the galactic gravitational potential is not important (i.e., $x_i \lesssim 1$), the outflows are distinguished by the initial values of $t_c/t_f$: (1) If initially $t_c/t_f \lesssim 1$, the gas cools as soon as it streams out of the galaxies. (2) If initially $t_c/t_f \gtrsim 1$, gas first cools adiabatically, and radiative cooling becomes dominant later on. (3) Radiative cooling cannot become important if $t_c/t_f \gg 1$ (e.g., $t_c/t_f \gtrsim 40$ for a gas with an initial temperature $10^7$ K) because the cooling time becomes too long. In the last case, the outflows either become galactic winds or stay in the galaxy as a corona, depending on whether the initial temperature is above the escape temperature of the galaxy. If, however, gravity is strong (i.e., $x_i \gtrsim 1$), the gas is pulled back to the galaxy before it has an opportunity to cool radiatively, and the outflow results in a hot corona. It is important to note that $t_c/t_f \sim 1$ at the base radius may not be used as the sole criterion for the onset of the radiative cooling; adiabatic expansion brings the gas temperature down, so $t_c/t_f$ decreases in the flow. Only when $t_c/t_f \gg 1$ can radiative cooling be ignored.

In case radiative cooling does become dominant, the gas temperature drops precipitously owing to the increase of the cooling rate with decreasing temperature. One may define a cooling radius where the gas temperature drops rapidly because of radiative cooling. Once radiative cooling sets in, the gas becomes unstable to convective and thermal instabilities, and the cloud formation ensues. The newly formed clouds inherit the kinetic energy of the gas at the point of cloud formation and coast farther from the cooling radius. For large initial temperature and shallow potential wells, the clouds escape the galaxy, while for galaxies with deep potentials, the clouds remain in the halo. Our general results on the outflows and the fate of the cooled gas are summarized in figure 2.

We apply our results to dwarf galaxies and normal galaxies. For reasonable ranges of the parameters, we find that dwarf galaxies generally have extended X-ray halos, and the outflows leave the galaxies eventually because of the relatively shallow potential wells. The gas may cool and forms clouds, in which case the cold clouds also leave the galaxy. If the gas density is low, radiative cooling never becomes important, in which case the outflow is in the form of galactic winds. In contrast, for large galaxies like our own, gravity is relatively strong, and the gas may not have a chance to cool before being pulled back to the galaxy. Even if the gas density is high enough so that radiative cooling becomes important eventually, the newly formed clouds are most likely to remain in the galaxy because of the strong gravitational field. In the extreme case where the temperature is larger than the escape temperature and radiative cooling is not important, galactic winds take place. We summarize the likely initial conditions for dwarf and normal galaxies in figure 2.

Some representative radial profiles of the gas temperature, the density, and the surface brightness of the cooling outflows are presented for dwarf and normal galaxies (figure 3). We find that the extent of the X-ray emission in the halo is usually much smaller than the outflow regions. The total radiation from the outflow that can be detected in the ROSAT band is about one order of magnitude smaller than the total radiation available for the initial temperature of about $3 \times 10^6$ K, and the fraction of the radiation emitted in the ROSAT band increases to about 30% of the total radiation if the initial temperature increases to $2 \times 10^7$ K. The X-ray in the hard (e.g., ASCA) energy band is smaller still, and we expect only the high temperature outflows ($T_i \gtrsim 10^7$ K) may be detected in the hard band. We calculate the mean temperature averaged over the entire outflow region, and we show that the mean temperature is usually a factor of 2-6 lower than the base temperature owing to the cooling of the gas away from the base (figure 4). This difference is important in understanding the physical conditions in the host galaxies, as usually only the mean gas temperature of the entire halo can be derived from observations. We also estimate the strength of the O VI emission lines ($\lambda = 1034$ Å). The mean surface brightness of the O VI doublets, especially from outflows in dwarf galaxies, is within the sensitivity of current UV experiments. We summarize the results of the calculations for the representative outflows in Table 1.

As mentioned in §1, the steady-outflow calculations may be applied to galaxies where the heating lasts longer than the flow time. However, for some extremely strong starburst galaxies, the flow region can be so large that the finite duration of the heating must be considered. An example is NGC 3628 ($v_{cir} = 240$ km/s) which has been detected by ROSAT to have an extended X-ray halo with radius of $\sim 30$ kpc (Dahlem et al. 1995). Since the galaxy may have experienced a starburst in the recent past (Fabbiano et al. 1990), the last period of heating by the burst of supernovae may be shorter than the crossing time for the outflow region. If so, time-dependent calculations are necessary. Also, since we have assumed spherical symmetry, our calculations cannot model nonradial outflows.

Our results may have significant implications for observations of QSO absorption line systems. More detailed discussion can be found in Wang (1995); here we only briefly mention possible applications of the model calculations presented above. Recent observations of the absorption lines in QSO spectra have increasingly shown evidence for the absorbing gas at large distances away from the associated galaxies, in particular for the heavy element absorbing gas, which has long been suspected to reside in galactic halos (Bahcall & Spitzer 1969). The identification of the galaxies causing the Mg II absorption lines in the intermediate redshifts show that the absorbing gas has a radius around the associated galaxies of about 50 kpc (e.g., Bergeron & Boisse 1991; Yanny et al. 1990; Bechtold & Ellingson 1992; Steidel 1993). Since the absorption lines of higher ionization states such as C IV are even more numerous (e.g., Sargent, Boksenberg, & Steidel 1988), this implies an even larger size for their absorbing regions if the gas is associated with galaxies. More recently, studies of the relation of the low column Lyman $\alpha$ forests with nearby galaxies show that they may be spatially correlated with the galaxies (Morris et al. 1993; Lanzetta et al. 1995), indicating a possible connection between galaxies and the absorption lines. If indeed Lyman $\alpha$ forests are also associated with galaxies, the size of the absorbing gas is even larger than that inferred for the heavy element absorbing gas, $\gtrsim 160$ kpc (Lanzetta et al. 1995). This size in fact agrees with the results of studies of Ly$\alpha$ lines in pairs of quasars at $z = 2$.

The huge size of the absorbing gas associated with the galaxies is rather hard to understand in the context of the normal galaxies like our own (Wang 1993). As seen in NGC 891 and from many other failed attempts to detect the halo X-ray emissions from normal galaxies, large halos are rare phenomena rather than the norm for normal galaxies. Indeed, estimates of the gas temperature of the hot phase in the Galactic ISM show that the temperature is $T_i \lesssim 10^6$ K (McKee & Ostriker 1978), and from figure 2 we see that the gas is trapped in the halo within a few kpc (also see Houck & Bregman 1990). Less massive galaxies, in particular dwarf galaxies, may, however, provide the needed gas at large distances away from galaxies. The gas temperature of the hot phase in less massive galaxies may be comparable to that of normal galaxies because the same heating mechanism through the shocks of supernova remnants is at work. But their gravitational potentials are much shallower than those of normal galaxies, therefore the gas is most likely to leave the galaxies. Our present work shows that the outflows from dwarf galaxies can cool radiatively for reasonable mass-loss rates (e.g., a few tenths of $M_\odot$/yr). The cooled gas is most likely to leave the galaxies as it inherits the kinetic energy of the hot gas which may not be exhausted by the cooling. The cooled gas thus can cause absorption over an area much larger than the original size of the galaxies. Note that radiative cooling of the gas may be crucial to accounting for both the observed low ionization states of the heavy elements and the narrow Ly$\alpha$ absorption lines.

I am especially grateful to Tim Heckman for useful discussions on observations of the gas outflows from galaxies throughout the course of the project. I also thank Mark Voit for helpful suggestions on the presentation of the present work. The author also thanks Steve Balbus, Mike Dahlem, and Bill Blair for useful discussions. Support from Alan C. Davis Fellowship at the Johns Hopkins University and Space Science Telescope Institute is gratefully acknowledged.

# APPENDIX
## STEADY, SPHERICAL FLOWS WITH CONSTANT $\gamma$

Here we consider the flow problem given by equations (2.1)-(2.3) but without radiative cooling ($A = 0$ or $q \to -\infty$). Unlike in the rest of the paper, the adiabatic index is taken as a constant that can be different from 5/3, emulating possible heating and cooling processes. Gravitational potential is assumed according to equation (2.9):

$$\phi(r) = -\frac{B}{nr^n}. \qquad (n \neq 0) \qquad (A1)$$

For $n = 0$, we assume a cut-off radius for the mass distribution in the galaxy, $r_{max}$, to avoid the logarithmic divergence, so within $r_{max}$ the potential is,

$$\phi(r) = -v_{cir}^2 \left[1 + \ln\left(\frac{r_{max}}{r}\right)\right], \qquad (n = 0) \qquad (A2)$$

where $v_{cir}$ is the circular (or rotation) velocity of the galaxy. Beyond $r_{max}$ the potential is the same as that of a point mass and is proportional to $1/r$ ($n = 1$ in eq. [A1]). Behaviors of the flow outside $r_{max}$ are readily understood from properties of the flow in a potential given by equation (A1) with $n = 1$, so for $n = 0$ we shall concentrate on studying the flow properties within $r_{max}$.

Flow solutions for spherical accretion problems with constant $\gamma$ in case $n = 1$ have been found by Bondi (1952), and the corresponding wind solutions have been studied by Parker (1958). A good review can be found in Holzer & Axford (1970). Here we extend flow solutions to arbitrary values of $n$ to take into account galactic gravitational potentials.

Upon taking $q \to -\infty$ in equations (2.6) and (2.7), we find that $M^2 = 1$ is a critical point, same as in the case with radiative cooling. To proceed smoothly through the sonic point, we require

$$2c^2 - r\frac{d\phi}{dr} = 0. \qquad (\text{at} \quad M^2 = 1) \qquad (A3)$$

The total energy per unit mass, $E$, is given by

$$E = \frac{v^2}{2} + \frac{\gamma}{\gamma - 1}\frac{p}{\rho} + \phi(r). \qquad (A4)$$

Substituting equation (A3) into (A4), for $n \neq 0$ we obtain

$$E = \frac{B}{nr_s^n}\left[\frac{(4+n) - (4-n)\gamma}{4(\gamma - 1)}\right], \qquad (A5)$$

where the subscript $s$ denotes the critical point where $M^2 = 1$ (sonic point). Namely, at the sonic point, the total energy is a critical fraction of the potential energy. To pass the sonic point smoothly, we require this fraction to be positive ($E > 0$). This is true only if

$$\gamma < \frac{4+n}{4-n} \leq \frac{5}{3}, \qquad (A6)$$

the last inequality is valid if $n \leq 1$. If $n = 1$, we obtain $\gamma < 5/3$, recovering the condition for transonic flows found by Bondi (1952) in accretion problems and by Parker (1958) in stellar winds. That is, for the point-mass potential, an adiabatic flow is either subsonic or supersonic. However, the condition for transonic flows is in general more stringent for potentials with exponent $n$ different from unity as readily seen in equation (A6). In fact, upon taking $n \to 0$, we find that the requirement becomes $\gamma < 1$. (The same result can be obtained by using equation [A2] directly). Obviously, flows cannot pass the sonic point smoothly even for an isothermal gas in a potential that gives rise to a flat rotation curve. The gravitational field in this case is much stronger than that of a point mass, and the nozzle effect necessary for a transonic flow is much more difficult to achieve. Thus, to overcome strong gravity, the gas has to be heated with increasing distance away from the origin in order to pass the sonic point.

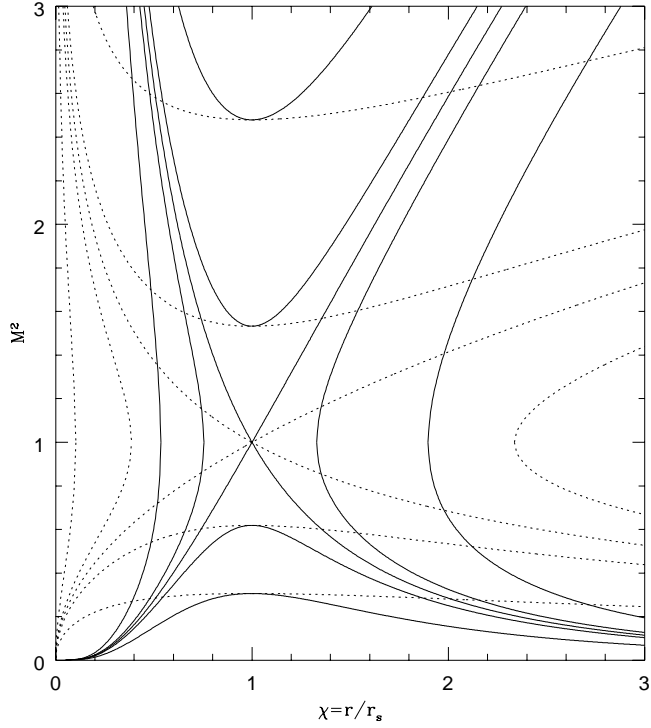

Fig. A1

Figure A1. *Mach number M as a function of the radius r normalized to the sonic radius $r_s$ (eq. [A10]). We take $\gamma = 1.25$. The solid curves are for the point-mass potential ($n = 1$) and the dotted curves are for $n = 0.5$ potential in equation (A1). The transonic solutions passing through the sonic point have $\delta = 1$ (eq. [A9]). The two supersonic (or subsonic) solutions for each n have $\delta = 0.99$ and $0.95$ away from the sonic point. The two solutions for each n at $\chi < 1$ (or $\chi > 1$) that proceed from supersonic to subsonic or vice versa but do not pass the sonic point have $\delta = 1.01$ and $1.05$ away from the sonic point (one of curves for $n = 0.5$ is out of the range to the right).*

To obtain flow solutions, we note that without radiative cooling, equations (2.1)-(2.3) can be integrated directly. Following Holzer & Axford (1970), we choose the mass-loss rate, $\lambda$, the energy per unit mass, $E$, and the specific entropy, $s = c_v \ln(P/\rho^\gamma)$ ($c_v$ is the specific heat capacity at a constant volume), as constants of integration.[3] The reference quantities are taken at the sonic point. From constant entropy, we may write the gas density

$$\rho = \gamma^{\frac{-1}{(\gamma-1)}} c^{\frac{2}{(\gamma-1)}} \exp\left[-\frac{s}{(\gamma-1)c_v}\right]. \tag{A7}$$

From the conservation of mass and using equation (A7), we obtain the sound speed in terms of the sound speed at the sonic point:

$$\left(\frac{c}{c_s}\right)^2 = \delta M^{-\frac{2(\gamma-1)}{(\gamma+1)}} \chi^{-\frac{4(\gamma-1)}{(\gamma+1)}}, \tag{A8}$$

where $\chi = r/r_s$, and

$$\delta = \exp\left[\frac{2(s-s_s)}{(\gamma+1)c_v}\right]. \tag{A9}$$

Again, subscript s denotes the variables at the sonic point. Substituting equation (A9) into equations (A4),

---

[3] An alternative approach is to use Bernoulli's equation directly (Bondi 1952).

and using equation (A5), we obtain solutions for *transonic flows* in terms of the Mach number for $n \neq 0$:

$$\frac{2}{n}\left[\frac{(4+n)-(4-n)\gamma}{4(\gamma-1)}+\chi^{-n}\right]\chi^{\frac{4(\gamma-1)}{(\gamma+1)}} = \delta\left(\frac{M^2}{2}+\frac{1}{\gamma-1}\right)M^{-\frac{2(\gamma-1)}{(\gamma+1)}}. \qquad (A10)$$

Note that solutions for $\gamma = 1$ can be obtained by taking the $(\gamma - 1)$ root on both sides of equation (A10); letting $\gamma \to 1$ and using the definition $(1 + \eta x)^{1/\eta} = e^x$ in the limit $\eta \to 0$, we have

$$\frac{4}{n\chi^n} + 4\ln\chi = M^2 - \ln M^2 + C_1, \qquad (A11)$$

where $C_1$ is a constant of integration.

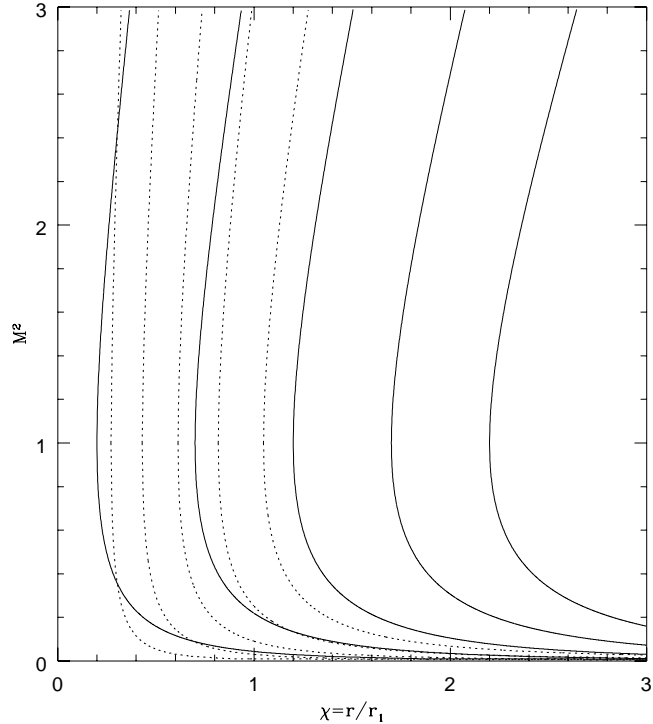

Fig. A2

Figure A2. Mach number $M$ for solutions with $E > 0$ as a function of the radius $r$ normalized to the chosen reference radius $r_1$ (see the text). The solid curves represent the flow solutions for the $n = 1$ potential in equation (A1) (the point-mass potential). We take $\gamma = 5/3$. The values of $\delta$ are from left to right 0.300, 0.425, 0.550, 0.675, and 0.800. The dotted curves denote the flow solutions with $E > 0$ in isothermal spheres (eq. [A2]), and we have taken $\chi_{max} = 3$. The values of $\delta$ are the same as those for the solid curves.

For isothermal spheres $n = 0$, similar transonic solutions can be obtained if the sonic point is inside $r_{max}$[4]:

$$2\left[\frac{(1+\gamma)}{4(\gamma-1)} - \ln\chi\right]\chi^{\frac{4(\gamma-1)}{(\gamma+1)}} = \delta\left(\frac{M^2}{2}+\frac{1}{\gamma-1}\right)M^{-\frac{2(\gamma-1)}{(\gamma+1)}}. \qquad (A12)$$

Some examples of transonic solutions (A10) with $n = 1$ and 0.5 are given in figure A1; transonic solutions (A12) for isothermal spheres with $\gamma < 1$ are similar, so we do not plot them separately here. Once the Mach number is known, it is straightforward to obtain the temperature ($\propto c^2$) and the density from equations (A8) and (A7).

---

[4] If the sonic point is outside $r_{max}$, the solutions are topologically similar to those in $n = 1$ potentials given by equation (A1) because outside $r_{max}$ gravitational field is equivalent to that by a point with a mass equal to the total mass within $r_{max}$.

If equations (A6) cannot be satisfied, transonic solutions are not possible. In this case, we may choose some different reference quantities other than at the sonic point, e.g., at $r = r_1$ (see below). From equation (A4), and using equations (A7) and (A8), we obtain for $n \neq 0$

$$\left(\frac{E}{v_0^2} + \chi^{-n}\right) \chi^{\frac{4(\gamma-1)}{(\gamma+1)}} = \delta \left(\frac{M^2}{2} + \frac{1}{\gamma-1}\right) M^{-\frac{2(\gamma-1)}{(\gamma+1)}}, \tag{A13}$$

where $v_0^2 = B/nr_1^n$. If $E > 0$, we choose $v_0^2 = E$, and if $E < 0$, we take $v_0^2 = -E$. If $E = 0$, $v_0^2$ can be chosen arbitrarily. Similar solutions can also be obtained for $n = 0$:

$$\left[\frac{E}{v_0^2} + 1 - \ln\left(\frac{\chi}{\chi_{max}}\right)\right] \chi^{\frac{4(\gamma-1)}{(\gamma+1)}} = \delta \left(\frac{M^2}{2} + \frac{1}{\gamma-1}\right) M^{-\frac{2(\gamma-1)}{(\gamma+1)}}, \tag{A14}$$

where $\chi_{max} = r_{max}/r_1$, $v_0^2 = v_{cir}^2$, and we have assumed $r_1 < r_{max}$. If $E > 0$, we choose $E = v_0^2$, and if $E < 0$, we choose $E = -v_0^2 \ln \chi_{max}$. Beyond $r_{max}$ (if solutions do exist), one may extend equation (A14) with solution (A13) with appropriate boundary conditions matched.

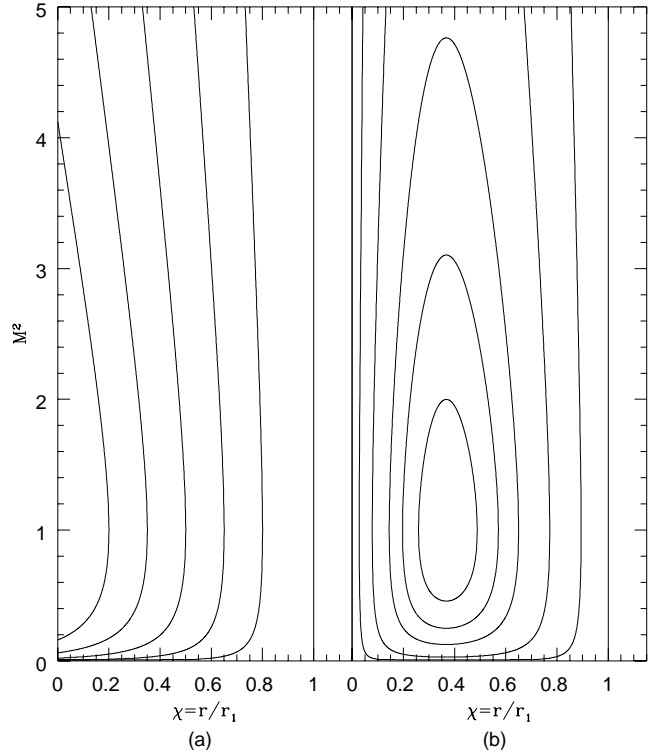

Fig. A3

Figure A3. a). Mach number $M$ for solutions with $E < 0$ as a function of the radius $r$ normalized to the maximum radius $r_1$ (vertical lines at $\chi = 1$) which the gas can reach. Gravitational potential is given by equation (A1) with $n = 1$ (the point-mass potential). We take $\gamma = 5/3$. The values of $\delta$ are from right to left 0.100, 0.175, 0.250, 0.325, and 0.400. b). Same as in a) except that the potential is given by equation (A2) (isothermal spheres). The values of $\delta$ from the outer to the inner curves are 0.050, 0.100, 0.140, 0.160, and 0.175.

Examples of the solutions for *either subsonic or supersonic* flow based on equations (A13) and (A14) with $E > 0$ are shown in figure A2. For $n = 0$, we have taken $\chi_{max} = 3$, but solutions for other values of $\chi_{max}$ are topologically similar. It is clear from equation (A13) that for the particular combination

$$n = \frac{(\gamma+1)}{4(\gamma-1)}, \tag{A15}$$

the Mach number is independent of radius for $E = 0$. For example, an adiabatic gas ($\gamma = 5/3$) has a constant Mach number in the point-mass potential ($n = 1$) if $E = 0$. Except for the special conditions imposed by equation (A15), the behavior of the solutions with $E = 0$ are similar to those with $E > 0$. However, solutions with $E < 0$ behave drastically differently from $E > 0$ solutions. It can be easily inferred from equations (A13) and (A14) that $r_1$ in this case defines a maximum radius beyond which flow solutions cannot exist. Physically, because the gas has an energy lower than that required to escape, it is trapped in the galaxy. In figure A3, we show some examples of solutions (A13) and (A14) for $E < 0$.

The asymptotic behavior of $M^2$ at $\chi \to \infty$ for both transonic flows and nontransonic flows with $E > 0$ can be obtained from equations (A10) and (A13). In the supersonic regime, we have

$$M^2 \to \chi^{2(\gamma-1)}, \tag{A16}$$

and in the subsonic regime we get

$$M^2 \to \chi^{-4}. \tag{A17}$$

From equations (A7) and (A8), it follows that the density in supersonic flows asymptotically approaches $\rho \propto \chi^{-2}$, while both the temperature and the density in subsonic flows approach a constant. Namely, a subsonic flow requires a confining pressure at infinity. Note that this asymptotic behavior can be also applied to isothermal spheres which have to be truncated at $r_{max}$; at $r \geq r_{max}$, the potential is of $n = 1$.

## Table 1

Model Calculations in §3b

|  |  | Dwarf Galaxies | | Normal Galaxies | |
|---|---|---|---|---|---|
| Input | $v_{cir}$ (km/s)[a] | 50 | 50 | 225 | 225 |
|  | $r_i$ (kpc)[b] | 0.3 | 0.3 | 3 | 3 |
|  | $T_i$ (K)[c] | $3 \times 10^6$ | $2 \times 10^7$ | $3 \times 10^6$ | $2 \times 10^7$ |
|  | $\lambda$ (M$_\odot$/yr)[d] | 0.3 | 10 | 3 | 100 |
| Result | $r_c$ (kpc)[e] | 1.5 | 5 | 13 | 50 |
|  | $v_c$ (km/s)[f] | 490 | 1310 | 295 | 1200 |
|  | $r_f$ (kpc)[g] | $\infty$ | $\infty$ | 32 | $\infty$ |
|  | L(total) (erg/s)[h] | $6 \times 10^{39}$ | $4 \times 10^{41}$ | $8 \times 10^{40}$ | $4 \times 10^{42}$ |
|  | L(0.1-2.2 keV) (erg/s)[i] | $4 \times 10^{38}$ | $1 \times 10^{41}$ | $6 \times 10^{39}$ | $1 \times 10^{42}$ |
|  | L(1.6-8.3 keV) (erg/s)[j] | $2 \times 10^{35}$ | $8 \times 10^{39}$ | $3 \times 10^{36}$ | $8 \times 10^{40}$ |
|  | L(O VI) (erg/s)[k] | $2 \times 10^{38}$ | $1 \times 10^{40}$ | $4 \times 10^{38}$ | $3 \times 10^{39}$ |
|  | F(O VI) (erg/s/cm$^2$/sr)[l] | $2 \times 10^{-7}$ | $1 \times 10^{-6}$ | $7 \times 10^{-9}$ | $4 \times 10^{-9}$ |

[a] Galaxy rotation velocity.
[b] The initial (base) radius.
[c] The initial temperature at the base.
[d] The mass-loss rate.
[e] The cooling radius.
[f] The velocity at the cooling radius, and thus the initial velocity of the clouds.
[g] The final radius which the cool clouds can coast to. The mass distribution is truncated at $r_{max} = 50 r_i$ (15 kpc and 150 kpc for dwarf and normal galaxies, respectively).
[h] The total luminosity integrated over the entire outflow.
[i] The luminosity in 0.1-2.2 keV band (corresponding to the ROSAT energy band) integrated over the entire outflow.
[j] The luminosity in 1.6-8.3 keV band (cf., the energy bands of GINGA, BBXRT and ASCA) integrated over the entire outflow.
[k] The total line luminosity from the O VI emission lines integrated over the outflow region inside $r_c$.
[l] The mean surface brightness of the O VI emission lines average over the outflow region inside $r_c$.